\documentclass[journal=jpcafh,manuscript=article]{achemso}

\usepackage[version=3]{mhchem} 
\usepackage{color}
\usepackage{float}
\usepackage{bm}
\usepackage{threeparttable}
\usepackage{amsmath,amsfonts,amssymb}
\usepackage{tablefootnote}
\usepackage{caption}
\usepackage{booktabs}

\newcommand{\cm}{cm$^{-1}$}

\usepackage{pdfpages}

\author{Saikiran Kotaru }
\affiliation{Department of Chemistry and Cherry L. Emerson Center for Scientific Computation, Emory University, Atlanta, Georgia 30322, U.S.A.}
\author{Chen Qu}
\email{szquchen@gmail.com}
\affiliation{Independent Researcher, Toronto, Ontario M9B0E3, Canada}

\author{Apurba Nandi}
\affiliation{Department of Chemistry and Cherry L. Emerson Center for Scientific Computation, Emory University, Atlanta, Georgia 30322, U.S.A.}
\author{Paul L. Houston}
\email{plh2@cornell.edu}
\affiliation{Department of Chemistry and Chemical Biology, Cornell University, Ithaca, New York
14853, U.S.A. }
\author{Joel M. Bowman}
\email{jmbowma@emory.edu}
\affiliation{Department of Chemistry and Cherry L. Emerson Center for Scientific Computation, Emory University, Atlanta, Georgia 30322, U.S.A.}

\title{ VPT2 Calculations of Vibrational Energies of \ce{CH3COOC6H4COOH} Done in Seconds on a Laptop Using a Machine Learned Potential }

\begin{document}








\newpage

\begin{abstract}
The determination of quartic force fields for use in vibrational second-order perturbation (VPT2) calculations, currently available in numerous electronic structure packages, becomes very expensive as the size of the molecule increases, especially if high-level coupled cluster theory is used. Machine-learned potentials (MLPs) for large molecules and clusters offer a viable alternative to obtain the quartic force field (QFF). Here, we report Fortran and Python software to determine the QFF and perform VPT2 calculations of energies from MLPs.  We describe this software briefly and then apply it to \ce{H2O} and protonated oxalate as test cases. The Fortran software is applied to 21-atom aspirin, using a fast MLP reported by us. Despite the fact that there are 32,509 unique cubic force constants for aspirin, the computer time to calculate them using this MLP is trivial, i.e., around one minute. These results are the first quantum anharmonic ones for such a large molecule. The present protocol offers an efficient way to study quantum anharmonic effects for vibrational energies in large molecules.  Currently, these are obtained overwhelmingly from classical molecular dynamics simulations, which cannot describe strong anharmonicity.
\end{abstract}

\newpage
Theoretical computation of vibrational energies is essential for interpreting spectroscopic measurements and is an active area of theoretical chemistry.\cite{bowman08, pccp13, jpca16, tenn16, carr17}
The simplest approaches, i.e., the harmonic approximation, assumes a quadratic separable potential and therefore ignores anharmonicity and mode coupling. Molecular dynamics, typically done at a fixed temperature of 300 - 500 K, is another widely-used especailly for large molecules.  This approach does not describe strong anharmonicity, especially for high-frequency X-H stretches, and mode coupling.
Vibrational second-order perturbation theory (VPT2)\cite{nielsen51} includes anharmonic effects through cubic and quartic force constants in a perturbative treatment, and thus achieves efficient computations compared to more rigorous variational approaches. However, the direct determination of these force constants in quartic force fields (QFFs), currently available in some electronic structure packages, becomes prohibitively expensive as the size of the molecule increases, especially if high-level coupled cluster energies are used.\cite{barone2026} To be specific, the number of unique cubic force constants increases as $N^3$ where $N$ is the number of vibrational normal modes. As noted in the Abstract, the number is 32,509 for 57-mode aspirin. These can be obtained using finite difference expressions using energies, gradients, or Hessians. For aspirin, using energies requires roughly 200,000 evaluations. Fewer terms are needed if gradients or Hessians are used, but these are more expensive to evaluate than energies (with the additional cost highly dependent on the level of \textit{ab initio} theory used). In short, a direct \textit{ab initio} calculation of the QFF for aspirin would be very computationally intensive. It should be noted, however, that efficient multi-level DFT approaches have been developed that permit VPT2 calculations with up to 50 atoms.\cite{barone2012toward,barone2026}

Machine-learned potentials for large molecules or clusters offer a viable alternative to obtain QFFs, among many other uses.  Recently, Meuwly and co-workers combined forces with software in Gaussian\cite{barone2004} to obtain the QFFs and perform subsequent VPT2 calculations, using PhysNet potentials for a number of molecules with up to ten atoms.\cite{VPT2Meuwly, VPT2FAD, Andreichev2025Dynamics}  Inspired by this work, we developed stand-alone software in Fortran and Python to determine the QFF using standard finite difference approaches using Hessians, gradients, or energies of MLPs. We then use Python software to perform the suite of VPT2 calculations directly using the QFF.  Here we demonstrate this for \ce{H2O} and protonated oxalate, as tests.  And then we apply the software to obtain VPT2 energies for the 21-atom aspirin. To the best of our knowledge, this is the largest molecule for which VPT2 energies have been reported, using an MLP. 

The paper is organized as follows. We first give a brief review of QFFs, VPT2 and the two widely used methods to deal with resonances, deperturbed VPT2 (DVPT2) and generalized VPT2 (GVPT2). The new Fortran and Python software to calculate the QFF from MLPs and then to obtain the VPT2, DVPT2, and GVPT2 energies is described.  Tests are presented for \ce{H2O}, protonated oxalate, and new calculations are reported for 21-atom aspirin (57 vibrational modes).

Recall that VPT2 is Rayleigh-Schr\"{o}dinger, second-order perturbation theory, where the separable harmonic-oscillator Hamiltonian (with neglect of vibrational angular momentum terms) is the zeroth-order model. In this approach, the potential, $V$, is expanded as a Taylor series about a minimum in terms of the (mass-scaled) normal coordinates, denoted $q_i$.\cite{spectro,barone2004,stanton2021} Truncation at the (separable) quadratic terms is the zeroth-order model, and the cubic and quartic terms of that series enter formally as terms of order $\lambda$ and $\lambda^2$, respectively.\cite{nielsen51} The QFF, in principle, contains all the cubic and quartic terms; however, for practical considerations, only the semi-diagonal quartic terms are retained. The cubic force constants are given by\cite{spectro, barone2004}
\begin{equation}
\phi_{ijk} = \frac{\partial^3 V}{\partial q_i \, \partial q_j \, \partial q_k},
\end{equation}
and the semi-diagonal quartic force constants are given by
\begin{equation}
\phi_{iijj} = \frac{\partial^4 V}{\partial q_i^2 \, \partial q_j^2},
\end{equation}
where $i, j, k$ label the vibrational normal modes.
As noted above, these force constants are generally obtained from direct electronic structure calculations numerically, and are by far the major source of the computational effort in VPT2 calculations.    

The VPT2 fundamental frequencies ($\nu_i$) for mode $i$ are given by\cite{spectro, barone2004}
\begin{equation}
\nu_i = \omega_i + 2\,\chi_{ii} + \frac{1}{2}\sum_{j \ne i} \chi_{ij}
,
\label{eq:vpt2}
\end{equation}
where $\omega_i$ is the harmonic vibrational frequency, and the anharmonicity constants $\chi_{ii}$ and $\chi_{ij}$ are given by
\begin{gather}
16\chi_{ii}
=
\phi_{iiii}
-
\sum_{j}
\frac{\left(8\omega_i^{2}-3\omega_j^{2}\right)\,\phi_{iij}^{\,2}}{\omega_j\left(4\omega_i^{2}-\omega_j^{2}\right)}
,
\label{eq:chi_ii} \\
4\chi_{ij}
=
\phi_{iijj}
-
\sum_{k}
\frac{\phi_{iik}\,\phi_{jjk}}{\omega_k}
+
\sum_{k}
\frac{2\omega_k\left(\omega_i^{2}+\omega_j^{2}-\omega_k^{2}\right)\,\phi_{ijk}^{\,2}}
{\Delta_{ijk}} 
+ 
\frac{4\left(\omega_i^{2}+\omega_j^{2}\right)}{\omega_i\,\omega_j}
\sum_{\tau}
B^{e}_{\tau}
\left(\zeta^{\tau}_{ij}\right)^2
,
\label{eq:chi_ij}
\end{gather}
where $B^{e}_{\tau}$ ($\tau=x,y,z$) are the equilibrium  rotational  constants, $\zeta^{\tau}_{ij}$ are the Coriolis coupling constants, and 
\begin{equation}
\Delta_{ijk}
=
(\omega_i+\omega_j+\omega_k)
(\omega_i+\omega_j-\omega_k)
(\omega_i-\omega_j+\omega_k)
(\omega_i-\omega_j-\omega_k).
\label{eq:delta}
\end{equation}
 
A well-known issue of this approach is evident from Eqs.~\eqref{eq:chi_ii} and \eqref{eq:chi_ij}, where the energy denominators approach zero in the presence of resonances, i.e., when $\omega_j \approx 2\omega_i$ (Fermi type-I) or $\omega_k \approx \omega_i + \omega_j$ (Fermi type-II). These near-degeneracies can lead to large errors in the computed vibrational energies. Methods to deal with resonances have been made and applied for at least thirty years and these are known as ``deperturbed VPT2'' (DVPT2) and ``generalized VPT2'' (GVPT2).\cite{martin1995, barone2004, barone2012toward, stanton2021, boyer2022wave, barone12, McCoy2022} These references also provide excellent background material on QFFs. The extensive developments of VPT2 and extensions to energies and properties\cite{barone2012toward} have made VPT2 a powerful and popular approach for anharmonic vibrational analysis.  Widely-used commercial and open-source electronic structure codes provide VPT2 as an option.  


As noted above, we have written stand-alone software in Fortran and Python to create a two-stage workflow. In the first stage, the QFF is calculated. In the Fortran implementation these force constants are obtained by finite differences of either energies or analytical gradients (which are preferred if they are available) using MLPs written in Fortran. In the Python implementation the QFF is obtained using finite difference expressions using gradients or Hessians. This software is tailored for MLPs written in Python. The second step uses Python software to read the QFF and to perform VPT2, DPVPT2 and GVPT2 calculations of anharmonic vibrational energies.  This software uses pieces of software contained in the PyVPT2.\cite{PyVPT2} This software makes direct calls to electronic energy software, e.g., PSI4, to obtain the QFF and then to perform the VPT2, DVPT2, and GVPT2 calculations.

Our software is first applied below for two tests; one on \ce{H2O} and the other on protonated oxalate anion. Then, we apply the software to obtain VPT2, DVPT2, and GVPT2 energies for the 57 fundamentals of the 21-atom aspirin using our MLP.\cite{pipaspirin} Comparisons with experiment (in solvent) are given. 

Tests for \ce{H2O} were done using the spectroscopically accurate potential of Partridge and Schwenke.\cite{Part97} This is a simple example, as there are no resonances for the fundamentals.  However, the effects of Coriolis coupling are significant owing to the large A-rotation constant.  The QFF was obtained using energies only and high-order finite difference approximations. VSCF/VCI calculations using MULTIMODE\cite{BowmanCarterHuang2003} were also performed and the results of the VPT2 and those calculation with and without Coriolis coupling are given in Table \ref{tab:coriolis_comparison}. As seen there is very good agreement with the VSCF/VCI results (denoted ``VCI" in the table).

\begin{table}[ht!]
\captionsetup[table]{justification=raggedright, singlelinecheck=false}
\centering
\caption{VPT2 and VCI fundamental energies (\cm) obtained with and without Coriolis coupling (CC) for water.}
\label{tab:coriolis_comparison}
\begin{tabular}{ccccc}

\hline
 Vibration Mode & VPT2 no CC & VPT2 with CC & VCI-no CC & VCI-with CC  \\
\hline
  bend & 1580.8 & 1594.2 & 1582.1 & 1594.0  \\
  symm stretch & 3654.4 & 3654.4 & 3656.9 & 3656.0   \\
  asymm stretch & 3739.7 & 3753.0 & 3742.8 & 3755.0 \\
\hline
\end{tabular}
\end{table}

Next, consider the 7-atom protonated oxalate anion where resonances play a role for some fundamentals; however, where Coriolis effects are very small and thus ignored, owing to the small rotation constants.  We first present results using a PhysNet neural network potential reported in Ref. \citenum{Andreichev2025Dynamics}. This  potential was initially trained with MP2/aug-cc-pVTZ energies and forces at 22,100 configurations and then transfer-learned with CCSD(T)/aug-cc-pVTZ energies and gradients at a subset of 2688 configurations. This potential is written in Python, with analytical Hessian available via auto-differentiation. The QFF was obtained using  finite differences applied to the Hessian.

In Table \ref{tab:oxa_vs_gaussian}, fundamentals from VPT2, DVPT2, and GVPT2 calculations using our  Python software are compared with those from reference \citenum{Andreichev2025Dynamics} which used Gaussian, with the same PhysNet MLP.
As seen, there are substantial differences between VPT2 and DVPT2 and GVPT2 results.  This is due to strong effects of resonances which are ignored in VPT2. Fundamentals using DVPT2 are in good agreement with the GVPT2 ones, but with 10--20 \cm ~discrepancy in certain modes. Finally our GVPT2 are within 3 \cm ~ or less of the GVPT2 results of ref. \citenum{Andreichev2025Dynamics}

\begin{table}[ht!]
\centering
\begin{threeparttable}
\begin{tabular}{rrrrr}
\hline
Number & VPT2 & DVPT2 & GVPT2 & GVPT2\tnote{a} \\
\hline
1  & 102.49  & 99.45   & 99.45  & 98.74  \\
2  & 291.55  & 291.45  & 291.46 & 288.14 \\
3  & 416.03  & 416.03  & 416.03 & 417.97 \\
4  & 479.55  & 478.40  & 478.40 & 474.20 \\
5  & 538.71  & 554.85  & 538.91 & 535.99 \\
6  & 690.99  & 690.88  & 690.89 & 691.15 \\
7  & 813.97  & 813.92  & 813.93 & 813.99 \\
8  & 840.47  & 840.40  & 840.41 & 838.94 \\
9  & 938.92  & 938.66  & 938.66 & 936.56 \\
10 & 1102.21 & 1096.00 & 1091.39 & 1089.69 \\
11 & 1277.33 & 1307.08 & 1302.46 & 1301.92 \\
12 & 951.343 & 1398.29 & 1383.62 & 1381.40 \\
13 & 1691.66 & 1691.37 & 1691.37 & 1695.42 \\
14 & 1837.74 & 1771.41 & 1763.27 & 1763.54 \\
15 & 2812.92 & 2765.26 & 2762.10 & 2767.01 \\
\hline
\end{tabular}
\begin{tablenotes}
    \item[a] Ref. \citenum{Andreichev2025Dynamics}
\end{tablenotes}
\end{threeparttable}
\caption{Fundamental energies (cm$^{-1}$) computed using VPT2, DVPT2 and GVPT2 for protonated oxalate using PhysNet potential. }
\label{tab:oxa_vs_gaussian}
\end{table}

\begin{figure}[ht!]
    \centering
    \includegraphics[width=1.1\linewidth]{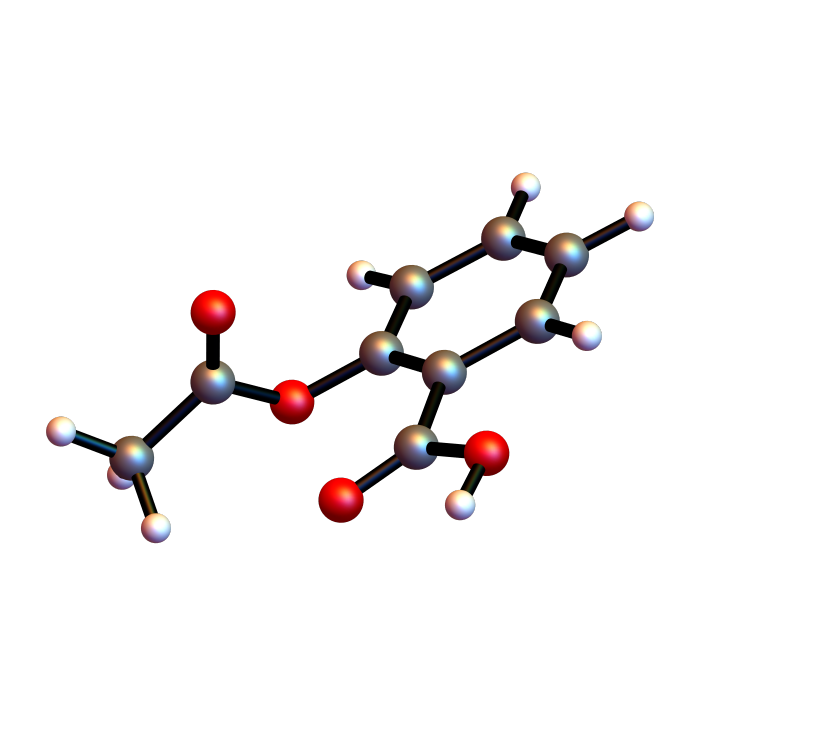}
    \caption{Equilibrium configuration of aspirin: white = H, red = O, gray = C.}
    \label{fig:aspirin}
\end{figure}

Given the good test results for \ce{H2O} and protonated oxalate, we applied our software to aspirin.
To obtain the QFF for aspirin, depicted in Figure \ref{fig:aspirin}, we used our MLP written in Fortran.\cite{pipaspirin} This MLP used permutationally invariant polynomial (PIP) regression to precisely fit energies and gradients from the rMD17 database for aspirin.\cite{Chmiela2017, Christensen2020} In brief, the potential surface is given  by
\begin{equation}
V(\bm{y})= \sum_{i=1}^{n_p} c_i p_i(\bm{y}),
\label{eq1}
\end{equation}
where $c_i$ are linear coefficients, $p_i$ are PIPs, $n_p$ is the total number of polynomials (and linear coefficients $c_i$) for a given maximum polynomial order, and $\bm{y}$ are transformed, Morse-like variables of internuclear distances $r_{ij}$ between atoms $i$ and $j$, i.e., $y_{ij}$=$\exp(-r_{ij}/a)$.\cite{Braams09} The coefficients $c_i$ were determined using a standard linear least-squares approach.

The dataset of energies and gradients for aspirin was obtained using NVT direct-dynamics calculations at 500 K and using PBE + TS-vdW\cite{TS-vdW} electronic structure method with the def2-SVP basis. The energy distribution for these geometries extends to almost 14,000 \cm.  

We reported several PIP PESs, and the largest and most precise fit is used here. This one contains 49,977 linear coefficients and is fit to 80,000 energies and 5,000 gradients for a total data size of 395,000.  The precision of the PES fit as compared to DFT values is given by eRMSE = 27 \cm ~and gRMSE = 53 \cm/bohr and by $R^2$ correlation coefficients of 0.999682 for energies and 0.999856 for gradients. Further details are given in ref. \citenum{pipaspirin}.

\newpage

Before presenting results of the VPT2 analysis for aspirin, we note how easily and fast it is to use VPT2 theory instead of the less accurate classical molecular dynamics approach. As noted in the title, it took less than one minute to calculated the QFF and perform the VPT2 analysis on a desktop using a single core of Intel i7-12700K CPU. Clearly, running classical MD for long enough time to get a reasonably well-resolved spectrum would take much more computational time. 

\begin{figure}[ht!]
    \centering
    \includegraphics[width=0.75\linewidth]{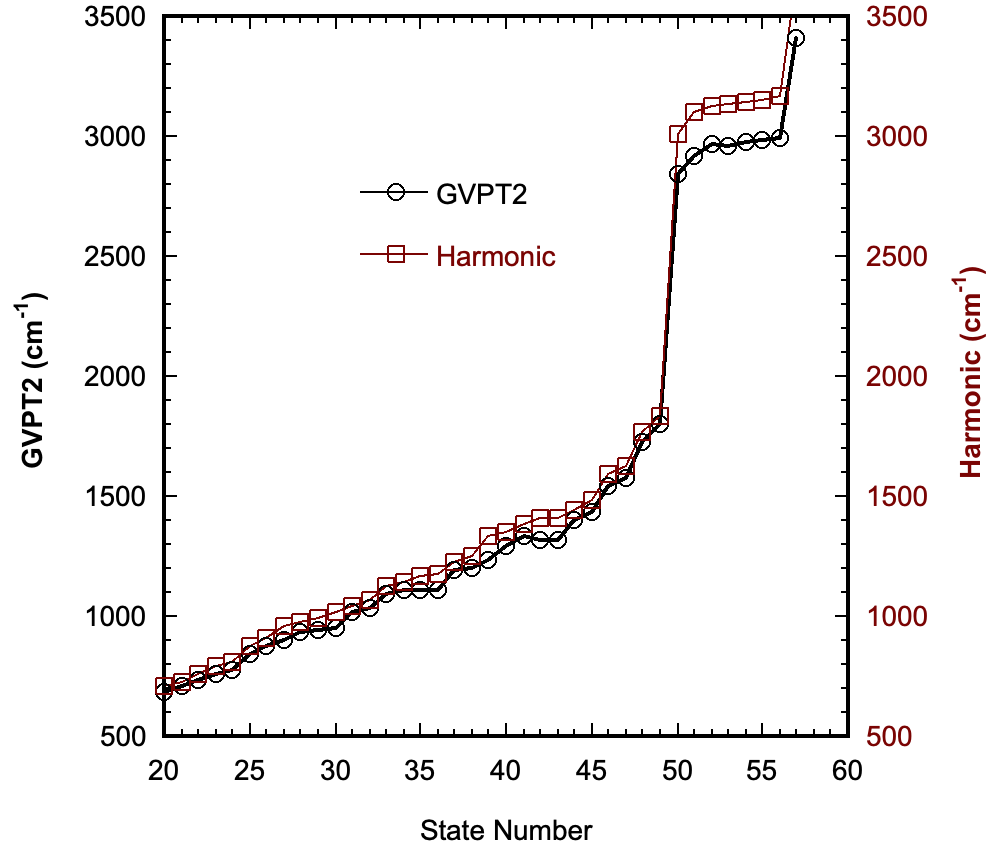}
    \caption{VPT2 and Harmonic Fundamental Energies of aspirin}
    \label{fig:VPT2HO}
\end{figure}

Energies of the 57 fundamentals from the GVPT2 and harmonic analysis are shown graphically in Figure \ref{fig:VPT2HO}. The numerical results and assignment of the higher-frequency fundamentals are given in the Supporting Information.  The VPT2 energies are all below the harmonic ones. The difference ranges from 1.9 \cm ~for the lowest harmonic frequency of 34 \cm, to 245 \cm ~for the highest harmonic frequency of 3652 \cm. And we believe the accuracy of the VPT2 energies are now mostly limited by the quality of the  electronic structure method used to obtain the data for the MLP.  

\begin{figure}[ht!]
    \centering
    \includegraphics[width=0.65\linewidth]{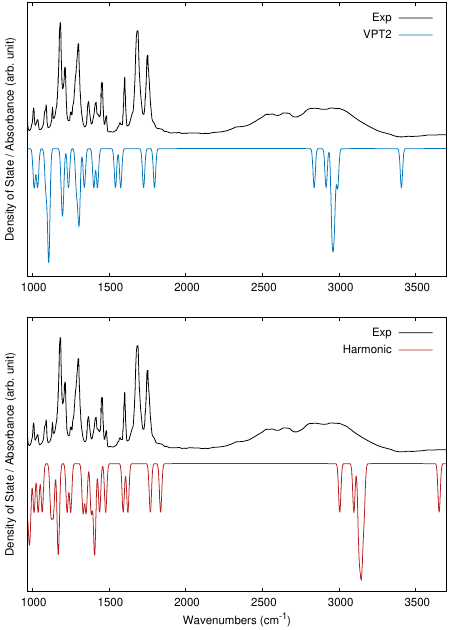}
    \caption{Experimental aspirin IR spectrum\cite{aspirinir} (black) and smooth density of states of VPT2 (blue) and Harmonic (red) of fundamental energies}
    \label{fig:spec}
\end{figure}

The IR spectrum of aspirin in condensed phase has been reported\cite{aspirinir}.  This spectrum is shown in Fig. \ref{fig:spec} along with the vibrational power spectrum from the GVPT2 and harmonic energies of the fundamentals. As seen, the GVPT2 spectrum aligns significantly better than the harmonic one with experiment, especially in the region of the broad higher frequency band between ca 2800 and 3200 \cm.  This is the region where GVPT2 fundamental energies are significantly below the harmonic ones by ca 150--200 \cm. The fundamentals in this band (see Table S1 in the SI) are seven CH-stretches from the methyl group and the ring.  Also, the gap in the IR spectrum between 2000 and 2500 \cm ~is reasonably reproduced by the GVPT2 power spectrum.

\noindent 

Typically, vibrational power spectra of large molecules using force fields or direct \textit{ab initio} calculations are obtained from Fourier transformation of a correlation function, e.g., the velocity autocorrelation function, from NVT molecular dynamics simulations.  These calculations are typically done at 300 K (roughly 200 \cm ~per vibrational degree of freedom).  Thus, for vibrational modes with frequencies above several hundred wavenumbers, especially stretches, the resulting spectra are essentially just the harmonic ones.  This is a well-known result. To account for anharmonicity, scaling factors, which have been used for many years, are sometimes applied.  For ``small'' molecules,  i.e., 10 atoms or so, quantum approaches, including second-order perturbation theory, multi-configuration time-dependent Hartree, and vibrational self-consistent field plus vibrational configuration interaction, are in widespread use.\cite{wsp} 

The present results are a major step in breaking the barrier to apply VPT2 to large molecules by using the recent advances in developing MLPs for such molecules. There are important caveats to note with this approach.  One is the ``no free lunch'' axiom.  While we have shown that the calculations of the QFF and VPT2 analysis is essentially trivial with an MLP, the cost of developing an MLP is not.  However, on the plus side, an MLP can be used for many studies in addition to VPT2 analysis, and so the cost is ``amortized'' over many uses beyond a QFF.  A second caveat is the intrinsic accuracy of an MLP.  There are two aspects here; one is the accuracy of the underlying electronic structure method used to obtain the data for the fit and the second is the precision error of the fit.  The second aspect is always addressed; however, results can vary significantly, depending on the ML method used.

To examine this sensitivity, we return to protonated oxalate for which a recent PIP PES, written in Fortran, was reported.\cite{QMchaos}  This PES was trained on the same datasets as the PhysNet potential; details of fit  are given in the SI of ref. \citenum{QMchaos} for the interested reader. We used this PES to obtain a QFF, using our Fortran software, with finite differences on analytical gradients as well as just energies. The resulting GVPT2 energies for the PIP PES are given in Table \ref{tab:compsox}, along with those using the PhysNet potential, given in Table 2.  As seen, with the exception of mode 15, results from PhysNet and PIP PES using gradients agree to within 0--3 \cm.  The mean absolute difference with GVPT2 energies using the PhysNet PES in the first column are also given. As seen, they are uniformly small, albeit largest for the QFF using finite difference on PIP energies, where the largest difference is 10 \cm ~for the lowest frequency mode.  Overall, the very good agreement between the PhysNet and PIP PESs is gratifying, as these PES are based on very different regression methods (see refs.\citenum{Andreichev2025Dynamics,QMchaos}) but produce high fit precision. Specifically, for CCSD(T) electronic energies up to 1000 \cm ~above the minimum, the
RMSEs are 0.39 \cm ~for the PhysNet PES and 0.49 \cm ~for the PIP PES.
Returning to the somewhat larger differences for mode 15, the OH-stretch, it has been noted  that this mode is very strongly coupled to other modes.\cite{wolke:2015,Andreichev2025Dynamics,QMchaos} In fact, VPT2 including GVPT2 breaks down for this mode, as the coupling is so strong such that this band is completely fractionated. So, it is likely that the 5-10 \cm~ differences are due to this strong coupling. A detailed examination of the QFFs might be very interesting for a future study. 

\begin{table}[H]
    \centering
    \caption{Comparison of GVPT2 fundamental energies of protonated oxalate from indicated sources and mean absolute difference (MAD) with respect to column one energies. Energies are in \cm.}
    \begin{threeparttable}
    \begin{tabular}{rccccc}
        \toprule
        & PhysNet\tnote{a} & PhysNet/G\tnote{b} & PIP\_PES\_g\tnote{c} & PIP\_PES\_e\tnote{d} \\
        \midrule
        1  &  99.5  &  98.7  &  99.9  & 110.0\\
        2  & 291.5  & 288.1  & 289.8  & 285.4  \\
        3  & 416.0  & 418.0  & 413.8  & 420.6  \\
        4  & 478.4  & 474.2  & 478.3  & 473.2  \\
        5  & 538.9  & 536.0  & 539.3  & 538.4  \\
        6  & 690.9  & 691.2  & 691.1  & 692.9  \\
        7  & 813.9  & 814.0  & 816.6  & 818.9  \\
        8  & 840.4  & 838.9  & 838.1  & 837.4  \\
        9  & 938.7  & 936.8  & 939.1  & 940.3  \\
        10 & 1091.4 & 1089.7 & 1089.4 & 1085.1  \\
        11 & 1302.5 & 1301.9 & 1304.3 & 1317.7  \\
        12 & 1383.6 & 1381.4 & 1383.2 & 1381.6 \\
        13 & 1691.4 & 1695.4 & 1691.7 & 1692.6  \\
        14 & 1763.3 & 1763.5 & 1763.5 & 1762.6  \\
        15 & 2762.1 & 2767.0 & 2756.2 & 2756.8  \\
        \midrule
        MAD & - & 2.1 & 1.4 & 4.6 \\
        \bottomrule
    \end{tabular}
    \begin{tablenotes}
        \item[a] PhysNet PES using our software;
        \item[b] PhysNet PES using Gaussian software as described in ref. \citenum{Andreichev2025Dynamics};
        \item[c] PIP PES using our software and finite difference of analytical gradients;
        \item[d] PIP PES our software and finite difference of energies.
    \end{tablenotes}
    \end{threeparttable}
    \label{tab:compsox}
\end{table}

The tentative conclusion from this study is that MLPs trained precisely on a dataset of including high energies may produce quantitatively useful QFFs; however, for states with strong resonance interactions, the results probably should be viewed with some caution.  As noted above for the aspirin dataset, the lower energies provided in the rMD17 dataset may actually be advantageous for QFF calculations.  (This observation is an updated and positive observation about this dataset, which we earlier critiqued as being too limited for some uses.\cite{MD17Pers})  
 
We conclude this section we note that the new software can also be used directly on aspirational universal force fields, such as MACE,\cite{mace2025} SO3LR,\cite{SOLR2025} and UMA,\cite{UMA} with the caveats about accuracy and precision kept in mind. And because the calculation of the QFF and subsequent VPT2 analysis is very fast given an MLP, it's clear that the speed of evaluation of energies and gradients from any MLP is essentially irrelevant. Also, beyond analysis at the minimum for anharmonic energies, we note that VPT2 theory can also be used at saddle points to obtain semi-classical tunneling corrections to standard transition state theory.\cite{sctst} This can also be done now for MLPs of large molecules, with the usual caveat about the accuracy of saddle points on those potentials. Finally, the results here for aspirin may also serve as benchmarks for approximate VPT2 approaches for larger molecules. For example, using a reduced number of modes, spectator modes, etc. as discussed in refs. \citenum{barone2026,fuse2024scaling}, or a fragmented local monomer approximation.\cite{YuVPT2} Finally, it is worth noting much earlier and related work in 1985 by Harding and Ermler.  They reported a Fortran code ``SURVIB''\cite{survib} in which a very precise fit is made using to a grid of electronic energies at configurations very close to the minimum. The fit is then used in  a normal mode analysis, and  elements of a QFF are obtained using finite-difference expressions for cubic and quartic force constants. These are then used in standard VPT2 calculations.  The code was demonstrated for formaldehyde using a low level of \textit{ab initio} theory, by current standards. Forty years later we are in the era of MLPs for molecules much larger than formaldehyde, but the software we report here is ``ancestrally" related to this early, pioneering work. 

To summarize, we reported software to obtain a quartic force field from general machine-learned potentials and used in VPT2 calculations of vibrational energies was reported  and tested on \ce{H2O} and the protonated oxalate anion. It was then applied to 21-atom aspirin using a previous PIP potential.  With this potential in hand the calculation of the QFF and VPT2 energies took roughly one minute of cpu time on a desktop computer.  Differences with the harmonic energies for 57 fundamentals range from 2 to 245 \cm.  Comparison with the only available low-resolution solution phase IR spectrum shows major improvement of the GVPT2 energies compared to the harmonic ones.  

The new software, written in Fortran and Python is available on request to the authors and will eventually be made available on Github. The QFF for aspirin is also available on request to the authors.





\begin{acknowledgement}

%
SK and JMB acknowledge support from NASA grant (80NSSC20K0360). We also thank Markus Meuwly, Valerii Andreichev and Silvan K{\"a}ser for discussions about their VPT2 calculations.

\end{acknowledgement}


\bibliography{refs}

@string(jpca="J. Phys. Chem. A")

@string(jcp="J. Chem. Phys.")

@string(cpl="Chem. Phys. Lett.")

@string(pccp="Phys. Chem. Chem. Phys.")

@string(ijqc="Int. J. Quant. Chem.")

@string(mp="Mol. Phys. ")

@string(jctc="J. Chem. Theory Comput.")

@string(jacs="J. Am. Chem. Soc.")

@article{nielsen51,
  title = {The Vibration-Rotation Energies of Molecules},
  author = {Nielsen, Harald H.},
  journal = {Rev. Mod. Phys.},
  volume = {23},
  issue = {2},
  pages = {90--136},
  year = {1951}
}

@article{martin1995,
  title={The anharmonic force field of ethylene, C2H4, by means of accurate ab initio calculations},
  author={Martin, Jan ML and Lee, Timothy J and Taylor, Peter R and Fran{\c{c}}ois, Jean-Pierre},
  journal=jcp,
  volume={103},
  number={7},
  pages={2589--2602},
  year={1995},
  publisher={American Institute of Physics}
}

@article{barone2004,
    author = {Barone, Vincenzo},
    title = {Anharmonic vibrational properties by a fully automated second-order perturbative approach},
    journal = jcp,
    volume = {122},
    number = {1},
    pages = {014108},
    year = {2004},
    month = {12},
}

@article{barone2012toward,
  title={Toward anharmonic computations of vibrational spectra for large molecular systems},
  author={Barone, Vincenzo and Biczysko, Malgorzata and Bloino, Julien and Borkowska-Panek, Monika and Carnimeo, Ivan and Panek, Pawel},
  journal={IJQC},
  volume={112},
  number={9},
  pages={2185--2200},
  year={2012},
  publisher={Wiley Online Library}
}

@article{fuse2024scaling,
  title={Scaling-up VPT2: A feasible route to include anharmonic correction on large molecules},
  author={Fus{\`e}, Marco and Mazzeo, Giuseppe and Longhi, Giovanna and Abbate, Sergio and Yang, Qin and Bloino, Julien},
  journal={Spectrochim. Acta A},
  volume={311},
  pages={123969},
  year={2024},
  publisher={Elsevier}
}

@article{stanton2021,
author = {Franke, Peter R. and Stanton, John F. and Douberly, Gary E.},
title = {How to VPT2: Accurate and Intuitive Simulations of CH Stretching Infrared Spectra Using VPT2+K with Large Effective Hamiltonian Resonance Treatments},
journal = jpca,
volume = {125},
number = {6},
pages = {1301-1324},
year = {2021},
}

@article{McCoy2022,
    author = {Boyer, Mark A. and McCoy, Anne B.},
    title = {A flexible approach to vibrational perturbation theory using sparse matrix methods},
    journal = jcp,
    volume = {156},
    number = {5},
    pages = {054107},
    year = {2022},
    month = {02},
}

@article{boyer2022wave,
  title={A wave function correction-based approach to the identification of resonances for vibrational perturbation theory},
  author={Boyer, Mark A and McCoy, Anne B},
  journal=jcp,
  volume={157},
  number={16},
  year={2022},
  publisher={AIP Publishing}
}

@INPROCEEDINGS{spectro,
   author = {{Gaw}, J.~F. and {Willetts}, A. and {Green}, W.~H. and {Handy}, N.~C.
	},
    title = "{Spectro: a Program for the Derivation of Spectroscopic Constants from Provided Quartic Force Fields and Cubic Dipole Fields}",
booktitle = {Advances in Molecular Vibrations and Collision Dynamics},
     year = 1991,
   editor = {{Bowman}, J.~M. and {Ratner}, M.~A.},
    pages = {169},
}

@article{PyVPT2,
    author = {Nelson, Philip M. and Sherrill, C. David},
    title = {PyVPT2: Interoperable software for anharmonic vibrational frequency calculations},
    journal = jcp,
    volume = {162},
    number = {3},
    pages = {032501},
    year = {2025},
    month = {01},
}

@article{barone12,
author = {Bloino,Julien  and Barone,Vincenzo },
title = {A Second-order Perturbation Theory Route to Vibrational Averages and Transition Properties of Molecules: General Formulation and Application to Infrared and Vibrational Circular Dichroism Spectroscopies},
journal = {J. Chem. Phys.},
volume = {136},
number = {12},
pages = {124108:1-15},
year = {2012}
}

@article{bowman08,
author = {Joel M. Bowman and Tucker Carrington  and Hans-Dieter Meyer },
title = {Variational Quantum Approaches for Computing Vibrational Energies of Polyatomic Molecules},
journal = {Mol. Phys.},
volume = {106},
number = {16-18},
pages = {2145--2182},
year = {2008}
}

@article{tenn16,
author = {Tennyson,Jonathan},
title = {Perspective: Accurate Ro-vibrational Calculations on Small Molecules},
journal = {J. Chem. Phys.},
volume = {145},
number = {12},
pages = {120901},
year = {2016},
doi = {10.1063/1.4962907}
}

@article{carr17,
author = {Carrington,Tucker},
title = {Perspective: Computing (Ro-)vibrational Spectra of Molecules with More than Four Atoms},
journal = {J. Chem. Phys.},
volume = {146},
number = {12},
pages = {120902},
year = {2017},
doi = {10.1063/1.4979117}
}

@article{wolke:2015,
  title={Diffuse vibrational signature of a single proton embedded in
                  the oxalate scaffold, HO$_2$CCO$_2^-$},
  author={Wolke, Conrad T and DeBlase, Andrew F and Leavitt,
                  Christopher M and McCoy, Anne B and Johnson, Mark A},
  journal=jpca,
  volume={119},
  number={52},
  pages={13018--13024},
  year={2015},
  publisher={ACS Publications}
}

@article{MD17Pers,
author = {Bowman,Joel M.  and Qu,Chen  and Conte,Riccardo  and Nandi,Apurba  and Houston,Paul L.  and Yu,Qi },
title = {The MD17 datasets from the perspective of datasets for gas-phase “small” molecule potentials},
journal = {J. Chem. Phys.},
volume = {156},
number = {24},
pages = {240901},
year = {2022},
doi = {10.1063/5.0089200},
}

@article{Part97,
author = {Partridge, Harry and Schwenke, David W.},
journal = {J. Chem. Phys.},
pages = {4618--4639},
title = {The Determination of an Accurate Isotope Dependent Potential Energy Surface for Water from Extensive ab Initio Calculations and Experimental Data},
volume = {106},
year = {1997}
}

@article{YuVPT2,
author = {Qi Yu and Joel M. Bowman},
title = {Vibrational second-order perturbation theory (VPT2) using local monomer normal modes},
journal = mp,
volume = {113},
number = {24},
pages = {3964--3971},
year = {2015},
publisher = {Taylor \& Francis},
}

@article{VPT2Meuwly,
author = {Käser, Silvan and Boittier, Eric D. and Upadhyay, Meenu and Meuwly, Markus},
title = {Transfer Learning to CCSD(T): Accurate Anharmonic Frequencies from Machine Learning Models},
journal = jctc,
volume = {17},
number = {6},
pages = {3687-3699},
year = {2021},
doi = {10.1021/acs.jctc.1c00249},
}

@Article{VPT2FAD,
author ="Käser, Silvan and Meuwly, Markus",
title  ="Transfer learned potential energy surfaces: accurate anharmonic vibrational dynamics and dissociation energies for the formic acid monomer and dimer",
journal  ="Phys. Chem. Chem. Phys.",
year  ="2022",
volume  ="24",
issue  ="9",
pages  ="5269-5281",
publisher  ="The Royal Society of Chemistry",
doi  ="10.1039/D1CP04393E",
url  ="http://dx.doi.org/10.1039/D1CP04393E"}

@misc{UMA,
      title={UMA: A Family of Universal Models for Atoms}, 
      author={Brandon M. Wood and Misko Dzamba and Xiang Fu and Meng Gao and Muhammed Shuaibi and Luis Barroso-Luque and Kareem Abdelmaqsoud and Vahe Gharakhanyan and John R. Kitchin and Daniel S. Levine and Kyle Michel and Anuroop Sriram and Taco Cohen and Abhishek Das and Ammar Rizvi and Sushree Jagriti Sahoo and Zachary W. Ulissi and C. Lawrence Zitnick},
      year={2026},
      eprint={2506.23971},
      archivePrefix={arXiv},
      primaryClass={cs.LG},
      url={https://arxiv.org/abs/2506.23971}, 
}

@book{wsp,
title = {Vibrational Dynamics of Molecules},
editor = {Bowman, Joel M},
publisher = {World Scientific},
year = {2022},
}

@article{Braams09,
author = {Braams, Bastiaan J. and Bowman, Joel M.},
title = {Permutationally Invariant Potential Energy Surfaces in High Dimensionality},
journal = {Int. Rev. Phys. Chem.},
volume = {28},
number = {4},
pages = {577--606},
year = {2009},
doi = {10.1080/01442350903234923}
}

@article{SOLR2025,
author = {Kabylda, Adil and Frank, J. Thorben and Suárez-Dou, Sergio and Khabibrakhmanov, Almaz and Medrano Sandonas, Leonardo and Unke, Oliver T. and Chmiela, Stefan and M{\"u}ller, Klaus-Robert and Tkatchenko, Alexandre},
title = {Molecular Simulations with a Pretrained Neural Network and Universal Pairwise Force Fields},
journal = jacs,
volume = {147},
number = {37},
pages = {33723-33734},
year = {2025},
doi = {10.1021/jacs.5c09558},
}

@article{mace2025,
author = {Kovács, Dávid P{\'e}ter and Moore, J. Harry and Browning, Nicholas J. and Batatia, Ilyes and Horton, Joshua T. and Pu, Yixuan and Kapil, Venkat and Witt, William C. and Magdău, Ioan-Bogdan and Cole, Daniel J. and Csányi, Gábor},
title = {MACE-OFF: Short-Range Transferable Machine Learning Force Fields for Organic Molecules},
journal = jacs,
volume = {147},
number = {21},
pages = {17598-17611},
year = {2025},
doi = {10.1021/jacs.4c07099},

}

@article{pipaspirin,
author = {Houston, Paul L. and Qu, Chen and Yu, Qi and Pandey, Priyanka and Conte, Riccardo and Nandi, Apurba and Bowman, Joel M.},
title = {No Headache for PIPs: A PIP Potential for Aspirin Runs Much Faster and with Similar Precision Than Other Machine-Learned Potentials},
journal = jctc,
volume = {20},
number = {8},
pages = {3008-3018},
year = {2024},
doi = {10.1021/acs.jctc.4c00054},
}

@Article{pccp13,
author ="Hochlaf, M.",
title  ="Editorial of the PCCP themed issue “Spectroscopy and dynamics of medium-sized molecules and clusters”",
journal  ="Phys. Chem. Chem. Phys.",
year  ="2013",
volume  ="15",
issue  ="25",
pages  ="9967--9969",
doi  ="10.1039/C3CP90047A",
url  ="http://dx.doi.org/10.1039/C3CP90047A"
}

@article{jpca16,
author = {Senent, Maria Luisa and Hochlaf, Majdi and Carvajal, Miguel},
title = {Spectroscopy and Dynamics of Medium-Sized Molecules and Clusters: Theory, Experiment, and Applications},
journal = jpca,
volume = {120},
number = {4},
pages = {475--476},
year = {2016},
doi = {10.1021/acs.jpca.5b12135},
}

@article{sctst,
title = {Semiclassical transition state theory. A new perspective},
journal = cpl,
volume = {214},
number = {2},
pages = {129-136},
year = {1993},
issn = {0009-2614},
doi = {https://doi.org/10.1016/0009-2614(93)90071-8},
author = {Rigoberto Hernandez and William H. Miller}
}

@article{survib,
author = {Harding, Lawrence B. and Ermler, Walter C.},
title = {Polyatomic, anharmonic, vibrational–rotational analysis. Application to accurate ab initio results for formaldehyde},
journal = {J. Comput. Chem.},
volume = {6},
number = {1},
pages = {13-27},
year = {1985}
}

@article{barone2026,
author = {Barone, Vincenzo and Lazzari, Federico and Mendolicchio, Marco},
title = {Accurate and Affordable Vibrational Spectra of Large Molecules: Primary, Auxiliary, and Spectator Modes in a Perturb-then-Diagonalize Framework},
journal = jctc,
volume = {22},
number = {5},
pages = {2453-2466},
year = {2026},
doi = {10.1021/acs.jctc.5c02123},
}

@article{BowmanCarterHuang2003,
   author = {Bowman, J. M and Carter, S. and Huang, X.},
   title = {MULTIMODE: a Code to Calculate Rovibrational Energies of Polyatomic Molecules},
   journal = {Int. Rev. Phys. Chem.},
   volume = {22},
   pages = {533--549},
   year = {2003},
   type = {Journal Article}
}

@article{Andreichev2025Dynamics,
   author = {Andreichev, Valerii and Käser, Silvan and Bocanegra, Erica L. and Salik, Madeeha and Johnson, Mark A. and Meuwly, Markus},
   title = {Dynamics of protonated oxalate from machine-learned simulations and experiment: infrared signatures, proton transfer dynamics and tunneling splittings},
   journal = pccp,
   volume = {27},
   number = {43},
   pages = {23288-23300},
   ISSN = {1463-9076},
   DOI = {10.1039/D5CP03085D},
   url = {http://dx.doi.org/10.1039/D5CP03085D},
   year = {2025},
   type = {Journal Article}
}

@article{Chmiela2017,
   author = {Chmiela, Stefan and Tkatchenko, Alexandre and Sauceda, Huziel E. and Poltavsky, Igor and Schütt, Kristof T. and Müller, Klaus-Robert},
   title = {Machine learning of accurate energy-conserving molecular force fields},
   journal = {Science Advances},
   volume = {3},
   number = {5},
   pages = {e1603015},
   DOI = {10.1126/sciadv.1603015},
   url = {https://doi.org/10.1126/sciadv.1603015},
   year = {2017},
   type = {Journal Article}
}

@article{Christensen2020,
doi = {10.1088/2632-2153/abba6f},
url = {https://dx.doi.org/10.1088/2632-2153/abba6f},
year = {2020},
month = {oct},
publisher = {IOP Publishing},
volume = {1},
number = {4},
pages = {045018},
author = {Anders S Christensen and O Anatole von Lilienfeld},
title = {On the role of gradients for machine learning of molecular energies and forces},
journal = {Machine Learning: Science and Technology},
}

@article{QMchaos,
    author = {Qu, Chen and Houston, Paul L. and Bowman, Joel M.},
    title = {Does the diffuse OH-stretch band in the IR spectrum of protonated oxalate exhibit quantum chaos?},
    journal = jcp,
    volume = {164},
    number = {6},
    pages = {061103},
    year = {2026},
    month = {02},
}

@misc{aspirinir,
  author       = {{Coblentz Society, Inc.}},
  title        = {Evaluated Infrared Reference Spectra},
  howpublished = {In NIST Chemistry WebBook, NIST Standard Reference Database Number 69},
  editor       = {P. J. Linstrom and W. G. Mallard},
  institution  = {National Institute of Standards and Technology},
  address      = {Gaithersburg, MD, USA},
  year         = {2026},
  doi          = {10.18434/T4D303},
  url          = {https://doi.org/10.18434/T4D303},
  note         = {Accessed: 2026-04-09}
}

@article{TS-vdW,
  title = {Accurate Molecular Van Der Waals Interactions from Ground-State Electron Density and Free-Atom Reference Data},
  author = {Tkatchenko, Alexandre and Scheffler, Matthias},
  journal = {Phys. Rev. Lett.},
  volume = {102},
  issue = {7},
  pages = {073005},
  numpages = {4},
  year = {2009},
  month = {Feb},
  publisher = {American Physical Society},
  doi = {10.1103/PhysRevLett.102.073005},
  url = {https://link.aps.org/doi/10.1103/PhysRevLett.102.073005}
}

\newpage
\includepdf[fitpaper=true, pages=-]{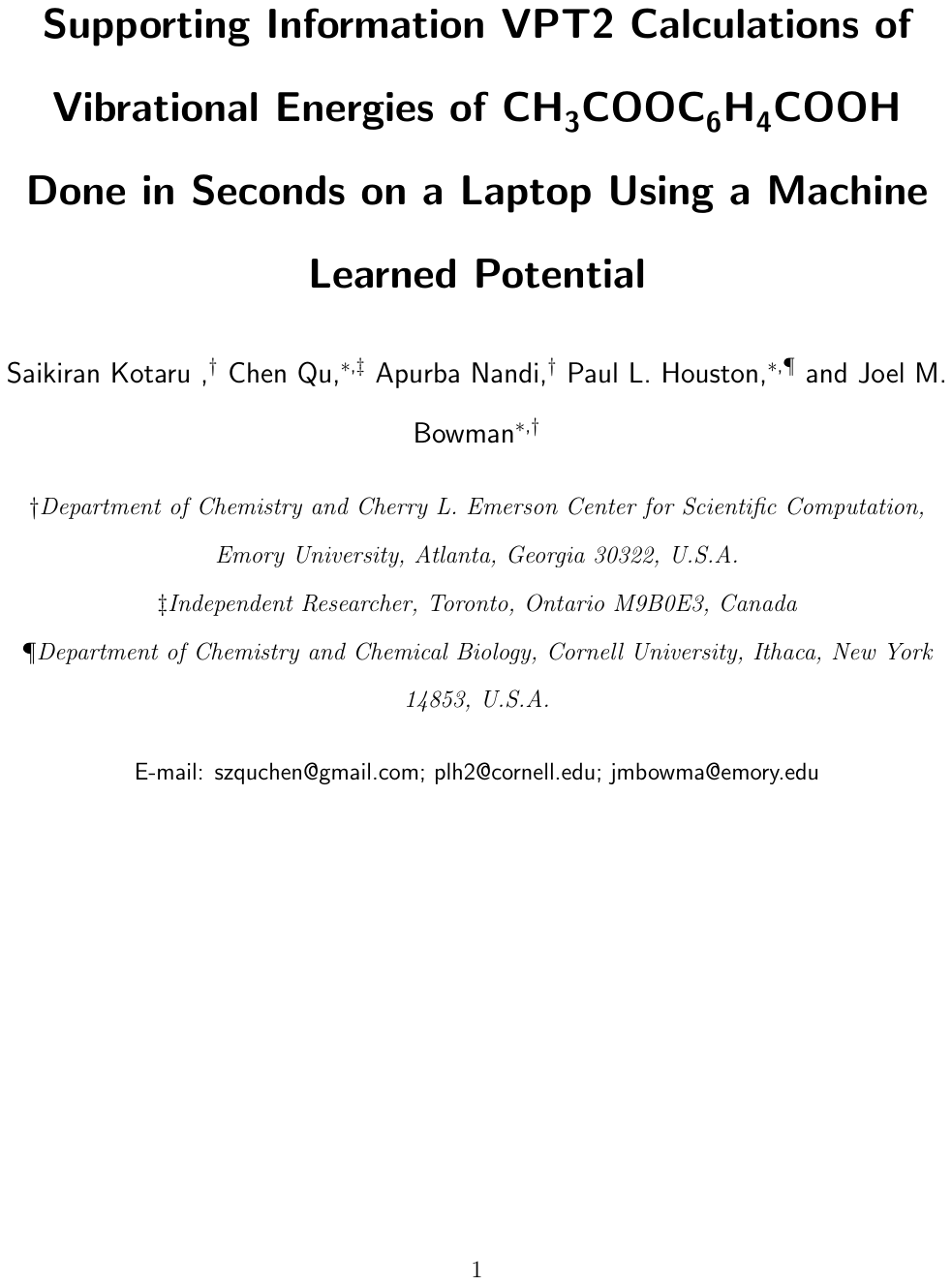}

\end{document}